\documentclass[aps,pra,reprint,showpacs,bibnotes,preprintnumbers,twoside,eqsecnum]{revtex4-1}

\usepackage{amsmath}
\usepackage{amssymb}
\usepackage{graphicx}
\usepackage{subfigure}
\usepackage{color}

\begin{document}


\title{Double-sided unidirectional reflectivity}
\author{J.~Ramirez-Hernandez}
\affiliation{Instituto de F\'{\i}sica, Benem\'erita Universidad Aut\'{o}noma de Puebla, Apdo. Post. J-48, Puebla, Pue., 72570, M\'{e}xico}

\author{F.~M.~Izrailev}
\email{felix.izrailev@gmail.com}
\affiliation{Instituto de F\'{\i}sica, Benem\'erita Universidad Aut\'{o}noma de Puebla, Apdo. Post. J-48, Puebla, Pue., 72570, M\'{e}xico}

\author{N.~M.~Makarov}
\email{makarov.n@gmail.com}
\affiliation{A.Ya.~Usikov Institute for Radiophysics and Electronics NASU, 61085 Kharkov, Ukraine}
\affiliation{Instituto de Ciencias, Benem\'erita Universidad Aut\'{o}noma de Puebla, Puebla, Pue., 72570, M\'{e}xico}

\date{\today}

\begin{abstract}
We study the effect of unidirectional reflectivity in the bilayer optical model with the balanced loss/gain terms. It was found that if the impedances characterizing the scattering part are different from those describing the leads, a new effect of double-sided unidirectional reflectivity emerges. In this case the vanishing of reflectivity occurs either for the left or right reflectivity, depending on the frequency of scattering wave. It is shown that the alternation of the unidirectional reflectivity from the left to the right can occur even if the corresponding Hamiltonian is non-$\mathcal{P}\mathcal{T}$-symmetric. Our analytical results are demonstrated by numerical data.
\end{abstract}


\maketitle

\section{Introduction}
Recent progress in the study of optical models with deliberately incorporated loss and gain has led to a new subject in optics that can be termed the $\mathcal{P}\mathcal{T}$-symmetric optics \cite{1,2,3,4,5,6,KLK12}. As is well known, the inclusion in the model either the loss or gain is directly related to non-Hermitian Hamiltonians corresponding to the scattering setup. Due to non-hermiticity its eigenvalues are, generically, complex, however, there is a specific situation when in spite of the non-Hermitian nature of the Hamiltonian, its eigenvalues can be real \cite{11,12}. This situation emerges when the Hamiltonian has the so-called $\mathcal{P}\mathcal{T}$-symmetry. Such a symmetry occurs if the Hamiltonian is symmetric under the combined action of space refection $\mathcal{P}$ and time reversal $\mathcal{T}$ operators. In this case the eigenvalues can be real in some frequency (energy) region.

In application to optics, the $\mathcal{P}\mathcal{T}$-symmetric scattering setup can be arranged \cite{3,4,Ro12} with the use of optical attenuation and amplifications in such a way that the complex refractive index distribution obeys the relation, $n(x) = n^{*} (-x)$ \cite{1,2}. This condition implies that real part of the refractive index is even function of position, $\mathrm{Re}\,n(-x)=\mathrm{Re}\,n(x)$, while its imaginary part is antisymmetric, $\mathrm{Im}\,n(-x)=-\mathrm{Im}\,n(x)$. In the scattering problem, one of the approaches is based on the transfer matrix $\hat{M}^{(T)}$ that allows one to describe all transport properties in dependence on the model parameters. Thus, the attention can be shifted from the symmetric properties of a Hamiltonian to those of the transfer matrix.

The eigenvalues of the transfer matrix that corresponds to the $\mathcal{P}\mathcal{T}$-symmetric Hamiltonian, undergo the transition from real to complex eigenvalues under the change of frequency. This effect is known in literature as "breaking the $\mathcal{P}\mathcal{T}$-symmetry". Note, however, that this term does not mean the breaking of any symmetry neither in the corresponding Hamiltonian nor in the transfer matrix $\hat{M}^{(T)}$. Actually, by "breaking the $\mathcal{P}\mathcal{T}$-symmetry" one means an existence of two regions in the frequency domain, with either real or complex eigenvalues of the transfer matrix. At the point of transition from one region to another it was found that the eigenvalues of the transfer matrix coincide; these points are known in the literature as the "exceptional points" (see, e.g. \cite{23} and references therein).

The interest in the $\mathcal{P}\mathcal{T}$-symmetric optical systems is due to anomalous properties of the transport characteristics, that are predicted and observed experimentally \cite{1,3,Ro12,Lo11,L09,RKEC10,VHIC14,VIC15}. One of such properties is the so-called {\it unidirectional reflectivity} emerging at specific values of the frequency. For this values one of the reflectances, either the left or right one, vanishes while the other remains finite. Typically, at such points the transmittance $T$ equals unity which is a generic property of the transfer matrix $\hat{M}^{(T)}$. If in addition the phase shift of the wave function after passing the scattering part is an integer modulo $2\pi$, in such a case one can speak of a non-visibility of the scattering setup.

In our recent paper \cite{RIMC16} it was found that for the chain of $N$ bilayers with the balanced loss/gain terms, the unidirectional reflectivity can emerge even when the scattering structure is a non-$\mathcal{P}\mathcal{T}$-symmetric.  In this case the frequencies for which one of the reflectances vanishes, are not those corresponding to the exceptional points. For this reason, below we refer to such points as the U-points, without any reference to the eigenvalues of the transfer matrix. Recently, the possibility to observe the real spectra in non-$\mathcal{P}\mathcal{T}$-symmetric models has been discussed in view of synthesizing complex potentials within the context of optical supersymmetry \cite{MHC13} (see, also, \cite{Correa}).

The goal of our present study is to demonstrate, both analytically and numerically, an emergence of the so-called {\it double-sided unidirectional reflectivity} that takes place in the scattering model consisting of two layers. In this model the impedances characterizing the scattering part are different from those describing the leads. Such a non-perfect matching has unexpectedly resulted in a new effect of the reflection properties. For the perfect matching with the leads (in the case of no loss/gain), the reflectance of the bilayer with balanced loss/gain vanishes at one side of the setup only, either the right or left one. We have found that the breaking of the perfect matching gives rise to an interchange of the left and right vanishing reflectivity when changing the wave frequency. Specifically, the left unidirectional reflectivity is transformed into the right one when the frequency crosses specific values. This effect depends entirely on the relation between the parameter characterizing the breaking of the perfect matching, and the parameter controlling the balanced gain and loss. This effect of the double-sided unidirectional reflectivity may be used in the experimental realizations of anomalous properties of scattering in optical devices.

\section{The model}

\begin{figure}[h!!!]
\centering
\includegraphics[scale=0.35]{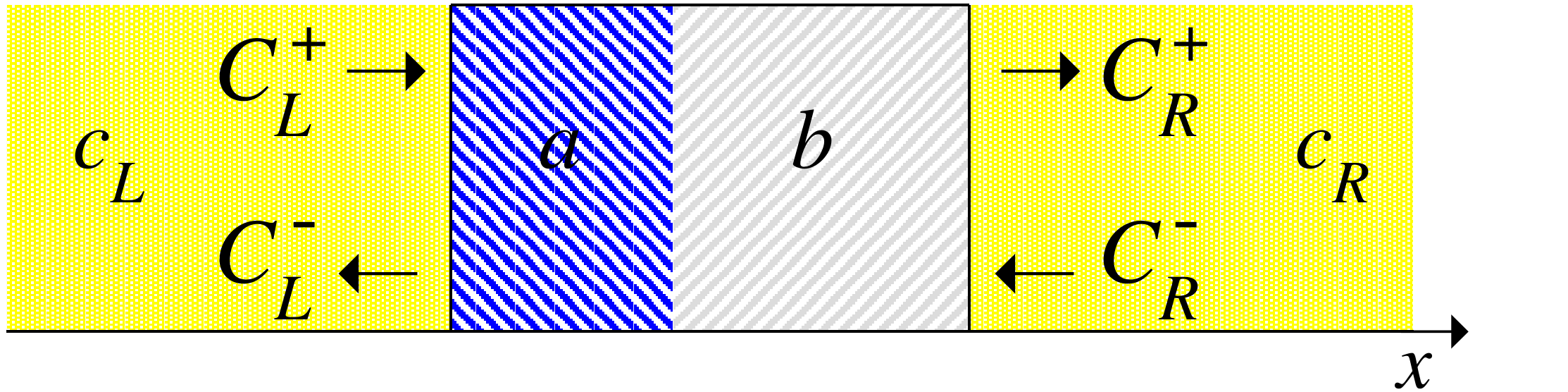}
\caption{(Color online) A sketch of the model.}\label{fig:Fig-01}
\end{figure}

The model describes the perpendicular propagation of an electromagnetic wave of frequency $\omega$ through a one-dimensional bilayer consisting of dielectric $a$ and $b$ slabs and connected to dielectric left, $c_L$, and right, $c_R$, perfect leads, see Fig.~\ref{fig:Fig-01}. The thicknesses of the layers are, respectively, $d_a$ and $d_b$, so that $d=d_a+d_b$ is the bilayer size. The $a$ and $b$ layers are made of the materials absorbing and amplifying the electromagnetic energy, respectively.
%
%
The loss and gain in the layers are incorporated via complex permittivities $\varepsilon_{a,b}$, while the magnetic permeabilities $\mu_{a,b}$ are assumed to be real and positive constants. The corresponding refractive indices $n_{a,b}=\sqrt{\varepsilon_{a,b}\,\mu_{a,b}}$, impedances $Z_{a,b}=\mu_{a,b}/n_{a,b}$ and wave phase shifts $\varphi_{a,b}=kn_{a,b}d_{a,b}$ of two constitutive $a$ and $b$ layers are chosen as ($k=\omega/c$),
\begin{subequations}\label{eq:BLMMDLG-nZphy}
\begin{eqnarray}
&&n_a=n_a^{(0)}(1+i\gamma),\qquad Z_a=Z(1+i\gamma)^{-1},\nonumber\\
&&\varphi_a=\frac{\varphi}{2}(1+i\gamma);\\
&&n_b=n_b^{(0)}(1-i\gamma),\qquad Z_b=Z(1-i\gamma)^{-1},\nonumber\\
&&\varphi_b=\frac{\varphi}{2}(1-i\gamma).
\end{eqnarray}
\end{subequations}
The dimensionless parameter $\gamma>0$ measures the stre\-ngth of balanced loss and gain inside $a$ and $b$ layers, respectively. The phase shift $\varphi=\varphi_a+\varphi_b>0$ of the wave passing the $(a,b)$ bilayer is real and positive.

Without loss/gain ($\gamma=0$) all the unperturbed optic parameters (refractive indices $n_{a,b}^{(0)}$, impedance $Z$, and phase shift $\varphi$) are real and positive constants. Moreover, as follows from the model \eqref{eq:BLMMDLG-nZphy}, we assume that in the case of no losses/gains the basic $a$ and $b$ slabs are perfectly matched and have the same phase shift $\varphi/2$,
\begin{subequations}\label{eq:MQS-def}
\begin{eqnarray}
&&Z=\mu_a/n_a^{(0)}=\mu_b/n_b^{(0)},\qquad\label{eq:MQS-Z}\\[6pt]
&&\varphi=2\omega n_a^{(0)}d_a/c=2\omega n_b^{(0)}d_b/c\,.\label{eq:MQS-phi}
\end{eqnarray}
\end{subequations}
This means that, correspondingly, their unperturbed impedances are equal, $\mu_a/n_a^{(0)}=\mu_b/n_b^{(0)}$, and their optic paths are the same, $n_a^{(0)}d_a=n_b^{(0)}d_b$. As for the leads, $c_L$ and $c_R$, only their impedances serve as necessary and sufficient optic characteristics. Here, we assume that the leads are prepared from the materials with the same impedance $Z_{c}=\mu_{c_L}/n_{c_L}=\mu_{c_R}/n_{c_R}$ (symmetric connection).

Within every $a$ or $b$ layer as well as inside the leads, the electric field $E(x)\exp(-i\omega t)$ of the propagating wave obeys the 1D Helmholtz equation. Its general solution inside every layer can be presented as a superposition of two plane waves traveling in both directions (see, e.g., Ref.~\cite{IKM12}). By combining these solutions with boundary conditions for the wave at the corresponding interfaces, one can obtain the relation that describes the wave transfer through the structure under consideration,
\begin{eqnarray}\label{eq:TotalTransfer}
&&\left(\begin{array}{c}C^{+}_{R}\\C^{-}_{R}\end{array}\right)=\hat{M}^{(T)}\left(\begin{array}{c}C^{+}_{L}\\C^{-}_{L}\end{array}\right),\nonumber\\[6pt]
&&\hat{M}^{(T)}=\hat{M}^{(\overrightarrow{ca})^{-1}}\hat{M}\hat{M}^{(\overrightarrow{ca})}.
\end{eqnarray}
This relation transforms the amplitudes of incident, $C^{+}_{L}$, and reflected, $C^{-}_{L}$, waves at the left side of the $(c_L|a)$ interface into the amplitudes of incident, $C^{-}_{R}$, and reflected, $C^{+}_{R}$, waves at the right side of the $(b|c_R)$ interface.

Matrix $\hat{M}^{(\overrightarrow{ca})}$ describes the wave transfer through the first interface from the left lead $c_L$ into slab $a$. It is defined by
\begin{equation}\label{eq:Qca-def}
\hat{M}^{(\overrightarrow{ca})}=\frac{1}{2}
\left(\begin{array}{ccc}1+Z_a/Z_c&&1-Z_a/Z_c\\[6pt]
1-Z_a/Z_c&&1+Z_a/Z_c\end{array}\right).
\end{equation}
Its inverse matrix $\hat{M}^{(\overrightarrow{ca})^{-1}}$ correspondingly reads
\begin{equation}\label{eq:Qac-def}
\hat{M}^{(\overrightarrow{ca})^{-1}}=\frac{1}{2}
\left(\begin{array}{ccc}1+Z_c/Z_a&&1-Z_c/Z_a\\[6pt]
1-Z_c/Z_a&&1+Z_c/Z_a\end{array}\right).
\end{equation}
Note that deriving the inverse matrix \eqref{eq:Qac-def} is equivalent to mutual (reciprocal) replacement of the indices in the initial matrix \eqref{eq:Qca-def}, i.e., $\hat{M}^{(\overrightarrow{ca})^{-1}}=\hat{M}^{(\overrightarrow{ac})}$. The determinants of the matrices are
\begin{equation}\label{eq:DetQcaac}
\det\hat{M}^{(\overrightarrow{ca})}=Z_a/Z_c,\qquad\det\hat{M}^{(\overrightarrow{ca})^{-1}}=Z_c/Z_a.
\end{equation}

It can be shown \cite{RIMC16} that the $(a,b)$ unit-cell transfer matrix $\hat{M}$ has the following elements,
\begin{eqnarray}\label{eq:BLMMDLG-M}
&&M_{11}=\frac{\exp(i\varphi)+\gamma^2\exp(-\gamma\varphi)}{1+\gamma^2},\nonumber\\
&&M_{12}=\frac{i\gamma}{1+\gamma^2}\left[\exp(-i\varphi)-\exp(\gamma\varphi)\right],\nonumber\\
&&M_{21}=\frac{i\gamma}{1+\gamma^2}\left[\exp(i\varphi)-\exp(-\gamma\varphi)\right],\nonumber\\
&&M_{22}=\frac{\exp(-i\varphi)+\gamma^2\exp(\gamma\varphi)}{1+\gamma^2}.
\end{eqnarray}
Note that the determinant of the matrix $\hat{M}$, as well as the determinant of the total transfer matrix $\hat{M}^{(T)}$, both are equal to unit due to the general condition inherent for transfer matrices,
\begin{subequations}\label{eq:DetMMT}
\begin{eqnarray}
\det\hat{M}&=&M_{11}M_{22}-M_{12}M_{21}=1,\\
\det\hat{M}^{(T)}&=&M_{11}^{(T)}M_{22}^{(T)}-M_{12}^{(T)}M_{21}^{(T)}=1.
\end{eqnarray}
\end{subequations}

The knowledge of the unit-cell transfer matrix $\hat{M}$ and the matrices \eqref{eq:Qca-def} -- \eqref{eq:Qac-def} describing the coupling to the left/right leads, allows one to obtain explicit expressions for the total matrix $\hat{M}^{(T)}$. As a result, the transmittance $T$, as well as, the left $R^{(L)}$ and right $R^{(R)}$ reflectances can be expressed according to their standard definitions,
\begin{eqnarray}
&&T=\left|M^{(T)}_{22}\right|^{-2}\nonumber\\
&&=\left[1+M^{(T)}_{12}M^{(T)}_{21}+M^{(T)}_{22}\left(M^{(T)*}_{22}-M^{(T)}_{11}\right)\right]^{-1},\qquad\label{eq:T-gen}\\[6pt]
&&\frac{R^{(L)}}{T} =\left|M^{(T)}_{21}\right|^2 ,\qquad\frac{R^{(R)}}{T} =\left|M^{(T)}_{12}\right|^2.\qquad\label{eq:R-gen}
\end{eqnarray}
The right part of Eq.~\eqref{eq:T-gen} directly follows from the unimodularity condition \eqref{eq:DetMMT} with the asterisk ``$*$" standing for complex conjugation.

It should be noted that the diagonal elements of $\hat{M}^{(T)}$ have the symmetry inherent for systems with the time-reversal symmetry,
\begin{equation}\label{eq:PatTimeRev}
M^{(T)}_{22}=M^{(T)*}_{11}.
\end{equation}
Therefore, all optical characteristics \eqref{eq:T-gen}, \eqref{eq:R-gen} of our system are specified only by off-diagonal elements of the total transfer matrix $\hat{M}^{(T)}$. As a result, the famous relation
\begin{equation}\label{eq:PatTimeRev-TR}
\left|1-T\right|=\sqrt{R^{(L)}R^{(R)}}
\end{equation}
holds that is widely discussed in view of $\mathcal{P}\mathcal{T}$-symmetric models (see, e.g., Ref.~\cite{GCS12}). However, the true time-reversal symmetry occurs when, in addition to Eq.~\eqref{eq:PatTimeRev} the off-diagonal elements of $\hat{M}^{(T)}$ meet the condition
\begin{equation}\label{eq:TRSym}
M^{(T)}_{21}=M^{(T)*}_{12}.
\end{equation}
In this case the transmittance $T\leqslant1$, the left/right reflectances become equal, $R^{(L)}=R^{(R)}=R$, and the relation \eqref{eq:PatTimeRev-TR} transforms to the flow conservation law,
\begin{equation}\label{eq:FlConsLaw}
T+R=1.
\end{equation}
As will be shown below, in our model with $\gamma\neq0$ the symmetry \eqref{eq:TRSym} is broken. Instead, another symmetry for off-diagonal elements of the total matrix $\hat{M}^{(T)}$ emerges.

\section{Perfect matching}

First, let us examine the situation when the leads are perfectly matched with $a$ and $b$ layers for $\gamma=0$,
\begin{equation}\label{eq:PMBLG-def}
Z_c=Z.
\end{equation}
In this case the mismatching ($Z_a\neq Z_c$ and $Z_b\neq Z_c$) emerge exclusively due to balanced loss/gain. Our analysis results in the following expressions for the matrix elements of $\hat{M}^{(T)}$,
\begin{eqnarray}\label{eq:PMBLG-Mtot}
M^{(T)}_{11}(\gamma)&=&\frac{\gamma^2\cosh(\gamma\varphi)+\exp(i\varphi)+i\gamma\sinh(\gamma\varphi)}{1+\gamma^2}\nonumber\\
&&-\frac{M^{(T)}_{21}(\gamma)-M^{(T)}_{12}(\gamma)}{2},\nonumber\\
M^{(T)}_{12}(\gamma)&=&\frac{i\gamma}{2(1+\gamma^2)}\,\mathcal{F}(-\gamma,\varphi),\\
M^{(T)}_{21}(\gamma)&=&\frac{i\gamma}{2(1+\gamma^2)}\,\mathcal{F}(\gamma,\varphi),\nonumber\\
M^{(T)}_{22}(\gamma)&=&\frac{\gamma^2\cosh(\gamma\varphi)+\exp(-i\varphi)-i\gamma\sinh(\gamma\varphi)}{1+\gamma^2}\nonumber\\[6pt]
&&+\frac{M^{(T)}_{21}(\gamma)-M^{(T)}_{12}(\gamma)}{2}.\nonumber
\end{eqnarray}
Here we have introduced the characteristic real-valued function,
\begin{equation}\label{eq:F-def}
\mathcal{F}(\gamma,\varphi)=(2+\gamma^2)\sinh(\gamma\varphi)-\gamma\sin\varphi+2[\cos\varphi-\cosh(\gamma\varphi)].
\end{equation}

As can be seen, the diagonal elements of $\hat{M}^{(T)}$ from Eqs.~\eqref{eq:PMBLG-Mtot} obey the symmetry \eqref{eq:PatTimeRev}, whereas the off-diagonal elements have a quite specific symmetry,
\begin{equation}\label{eq:BLG-Msym}
M^{(T)}_{22}(\gamma)=M^{(T)*}_{11}(\gamma),\quad M^{(T)}_{21}(\gamma)=M^{(T)*}_{12}(-\gamma).
\end{equation}
As is shown in Refs.~\cite{L10,M13,RIMC16,VHIC14,VIC15}, such a symmetry of the transfer matrix occurs in $\mathcal{P}\mathcal{T}$-symmetric models, however, can also emerge, as in our case, even if the corresponding Hamiltonian is not $\mathcal{P}\mathcal{T}$-symmetric.

Now, in accordance with definitions \eqref{eq:T-gen} and \eqref{eq:R-gen} the analytical expressions for the transmittance $T$, and left and right reflectances, $R^{(L)}$ and $R^{(R)}$, read
\begin{eqnarray}
&&T=\left[1-\frac{\gamma^2}{4(1+\gamma^2)^{2}}\,\mathcal{F}(\gamma,\varphi)\mathcal{F}(-\gamma,\varphi)\right]^{-1};\label{eq:PMBLG-T}\\
&&\frac{R^{(L)}}{T}=\frac{\gamma^2}{4(1+\gamma^2)^{2}}\,\mathcal{F}^2(\gamma,\varphi)\,;\label{eq:PMBLG-lR}\\
&&\frac{R^{(R)}}{T}=\frac{\gamma^2}{4(1+\gamma^2)^{2}}\,\mathcal{F}^2(-\gamma,\varphi)\,.\label{eq:PMBLG-rR}
\end{eqnarray}
As a consequence of the symmetry \eqref{eq:BLG-Msym}, see also Eqs.~\eqref{eq:PMBLG-Mtot}, the transmittance is an even function of the loss/gain parameter $\gamma$, while the reflectances transform into each other one as the sign before $\gamma$ changes,
\begin{equation}\label{eq:PMBLG-Teven-Rodd}
T(-\gamma)=T(\gamma),\qquad R^{(R)}(-\gamma)=R^{(L)}(\gamma).
\end{equation}
In addition, due to the partial time-reversal symmetry \eqref{eq:PatTimeRev}, the quantities $T(\gamma)$, $R^{(L)}(\gamma)$ and $R^{(R)}(\gamma)$ satisfy the relation \eqref{eq:PatTimeRev-TR}.

It is important to stress that the function $\mathcal{F}(-\gamma,\varphi)$ is negative ($\mathcal{F}(-\gamma,\varphi)<0$) for any values of $\gamma > 0$ and $\varphi>0$. Unlike this, the function $\mathcal{F}(\gamma,\varphi)$ can be either positive or negative as a function of $\varphi$. It is highly important that this function can vanish at specific points $\varphi=\varphi_s(\gamma)$. In what follows we term these points the {\it unidirectional points} (or, U-points) since at these points the left reflectance vanishes, however, the other one remains finite, see Eqs.~\eqref{eq:PMBLG-lR} and \eqref{eq:PMBLG-rR}.

\begin{figure}[!t]
\includegraphics[width=\columnwidth]{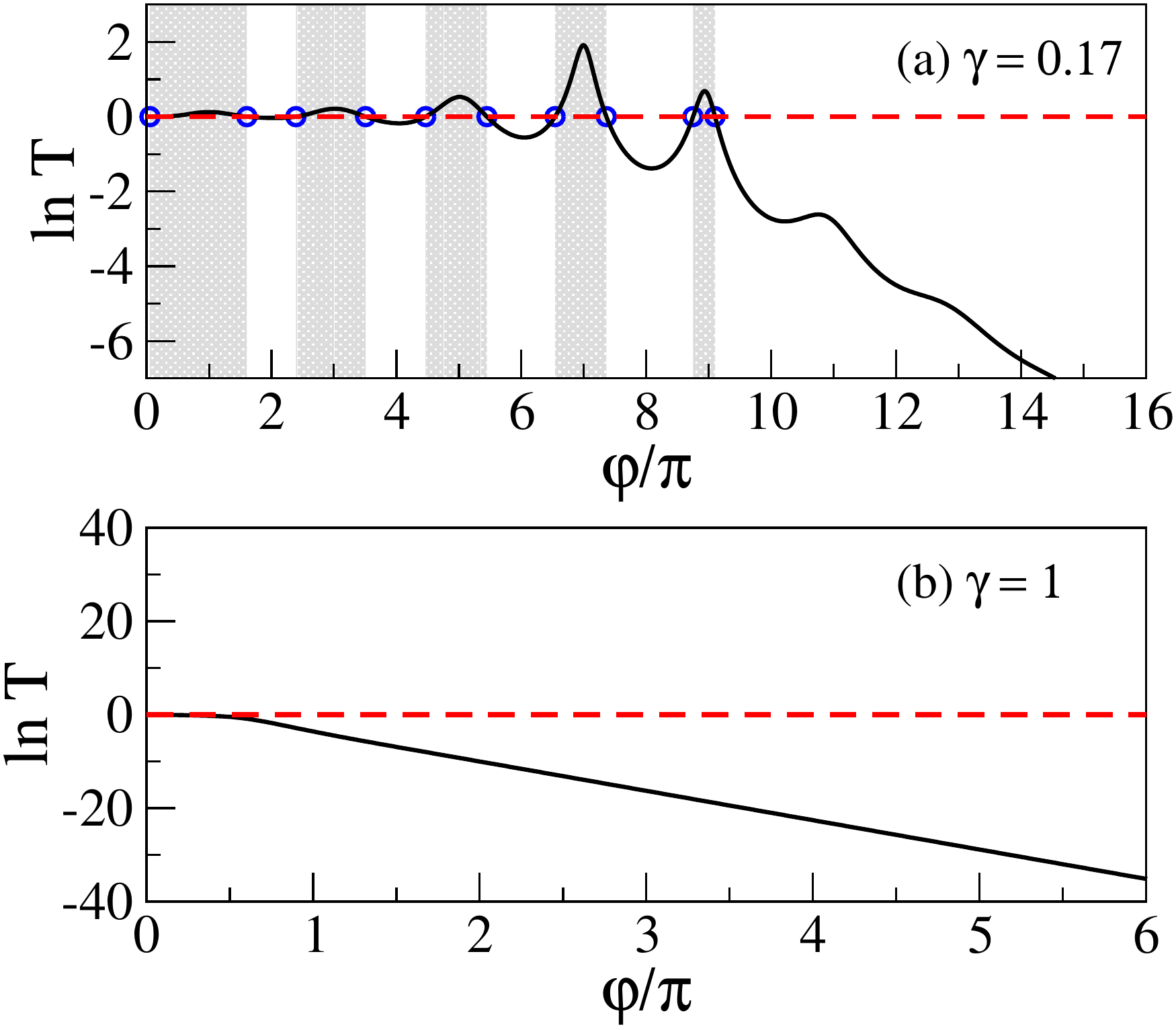}
\caption{The logarithm of $T$ defined by Eq.~\eqref{eq:PMBLG-T} as a function of $\varphi$ for $\gamma=0.17$ and $\gamma=1$. Shaded regions correspond to $T>1$, and circles stand for the borders (specific U-points $\varphi=\varphi_s$) between the regions with $T<1$ and $T>1$.}\label{fig:Fig-02}
\end{figure}

Evidently, for $\gamma=0$ the system is completely transparent, i.e. its transmittance equals unity and both reflectances vanish,
\begin{equation}\label{eq:PMBLG-TR-gamma0}
T=1,\qquad R^{(L)}=R^{(R)}=0\qquad\mathrm{for}\quad\gamma=0.
\end{equation}

\begin{figure}[t]
\centering
\includegraphics[width=\columnwidth]{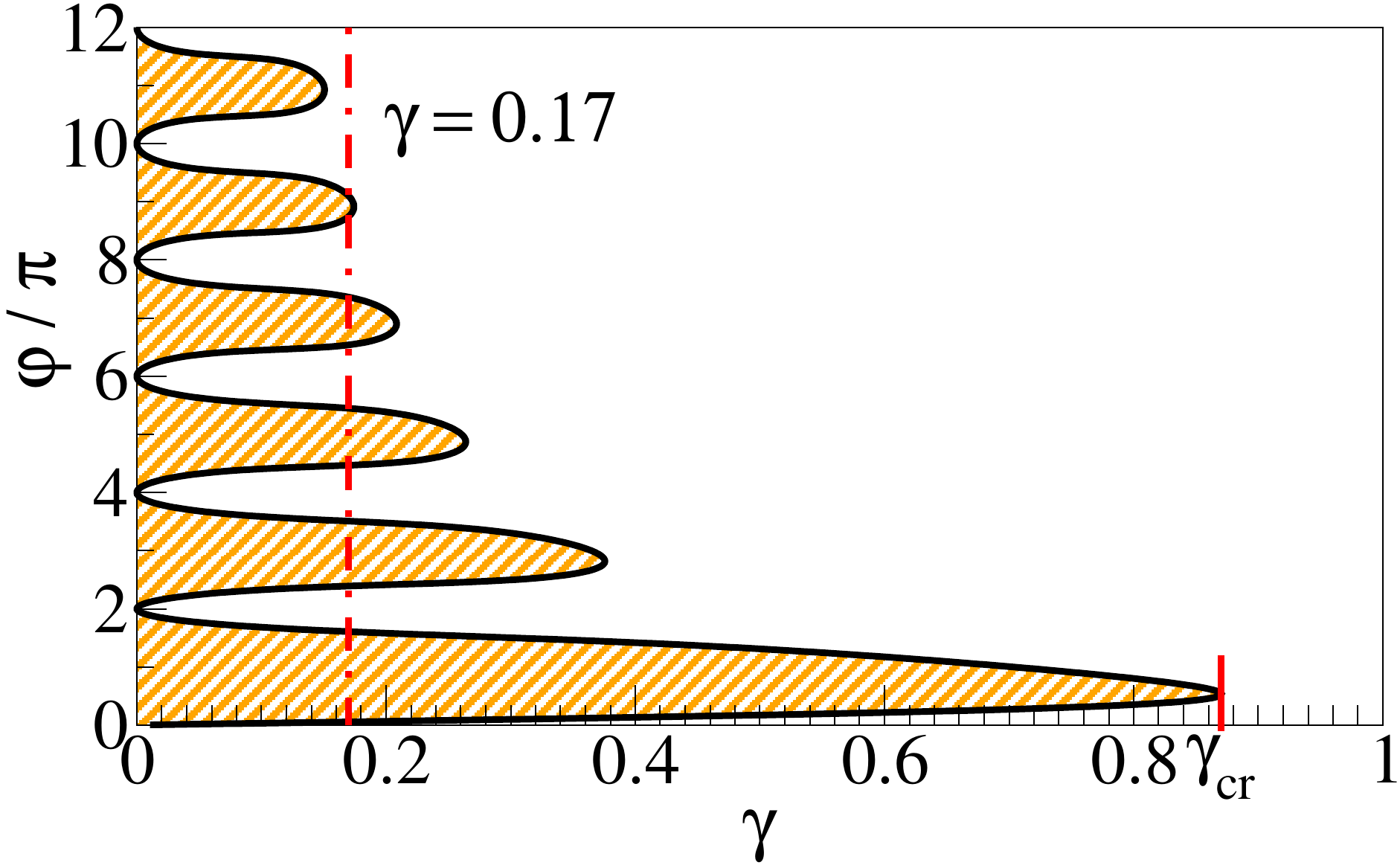}
\caption{The diagram $(\gamma,\varphi/\pi)$. Full curve shows the border $\varphi=\varphi_s(\gamma)$ on which $T=1$. To the right from it (blank space), $T(\varphi)<1$. Inside shaded regions $T(\varphi)>1$. The vertical dash-dotted line is shown for $\gamma=0.17$.}\label{fig:Fig-03}
\end{figure}

Two examples of the dependence $T(\varphi)$, see Eq.~\eqref{eq:PMBLG-T}, for $\gamma\neq0$ are given in Fig.~\ref{fig:Fig-02}. The analysis shows that a distinctive peculiarity of the transmittance is its oscillations around the value $T=1$, see Fig.~\ref{fig:Fig-02}~(a). The number of regions with $T>1$ and those with $T<1$ are finite and determined by the loss/gain parameter $\gamma$ only. At the borders between these regions denoted by circles in Fig.~\ref{fig:Fig-02}~(a), the transmission is perfect ($T=1$), and this happens due to vanishing the function $\mathcal{F}(\gamma,\varphi)$. Thus, the equation $\mathcal{F}(\gamma,\varphi)=0$ determines the values of the U-points $\varphi_s(\gamma)$ for which $T=1$, in dependence on the parameter $\gamma$,
\begin{equation}\label{eq:PMBLG-SpPoints}
\mathcal{F}(\gamma,\varphi)=0\,\to\quad\varphi=\varphi_s(\gamma)\,\to\quad T(\varphi_s)=1.
\end{equation}
The set of such U-points $\varphi_s(\gamma)$ is presented in Fig.~\ref{fig:Fig-03}. One can see that for a given value of $\gamma$ from the interval $0<\gamma\leqslant\gamma_{cr}\approx0.8703$, the number of specific U-points is finite. It can be shown that the smaller the value of $\gamma$ the larger the number of specific points $\varphi=\varphi_s(\gamma)$ and, correspondingly, the more the oscillations of the transmittance $T(\varphi)$.

As one can see in Fig.~\ref{fig:Fig-03}, the critical value $\gamma_{cr}$ is defined as the maximal value of $\gamma$ at which only one U-point $\varphi_s(\gamma_{cr})\approx0.53\pi$ exists, therefore, the equation \eqref{eq:PMBLG-SpPoints} still has a solution.

If $\gamma$ exceeds the critical value, $\gamma>\gamma_{cr}$, the function $\mathcal{F}(\gamma,\varphi)$ becomes positive, $\mathcal{F}(\gamma,\varphi)>0$, for any value of phase shift $\varphi>0$. Consequently, there are no U-points $\varphi_s(\gamma)$ and the regions with $T(\varphi)>1$ disappear. Here the transmittance decreases exponentially with an increase of $\varphi$ (or, the same, with the increase of $\omega$), see Fig.~\ref{fig:Fig-02}~(b).
\begin{figure}[!ht]
\centering
\includegraphics[width=\columnwidth]{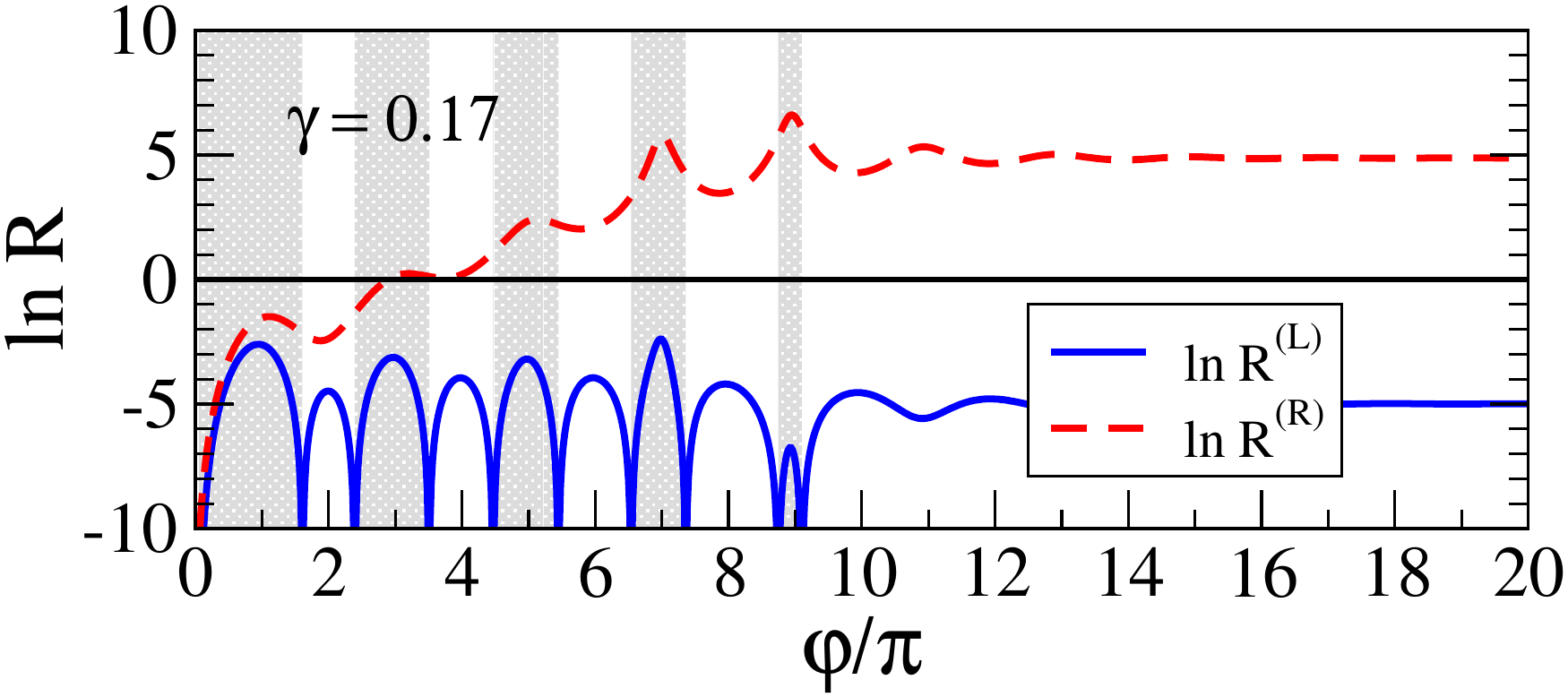}
\caption{The dependence of the left $R^{(L)}$ and right $R^{(R)}$ reflectances on the phase shift $\varphi$ (the wave frequency $\omega$) for $\gamma=0.17$.}\label{fig:Fig-04}
\end{figure}

\begin{figure}[!ht]
\centering
\includegraphics[scale=0.45]{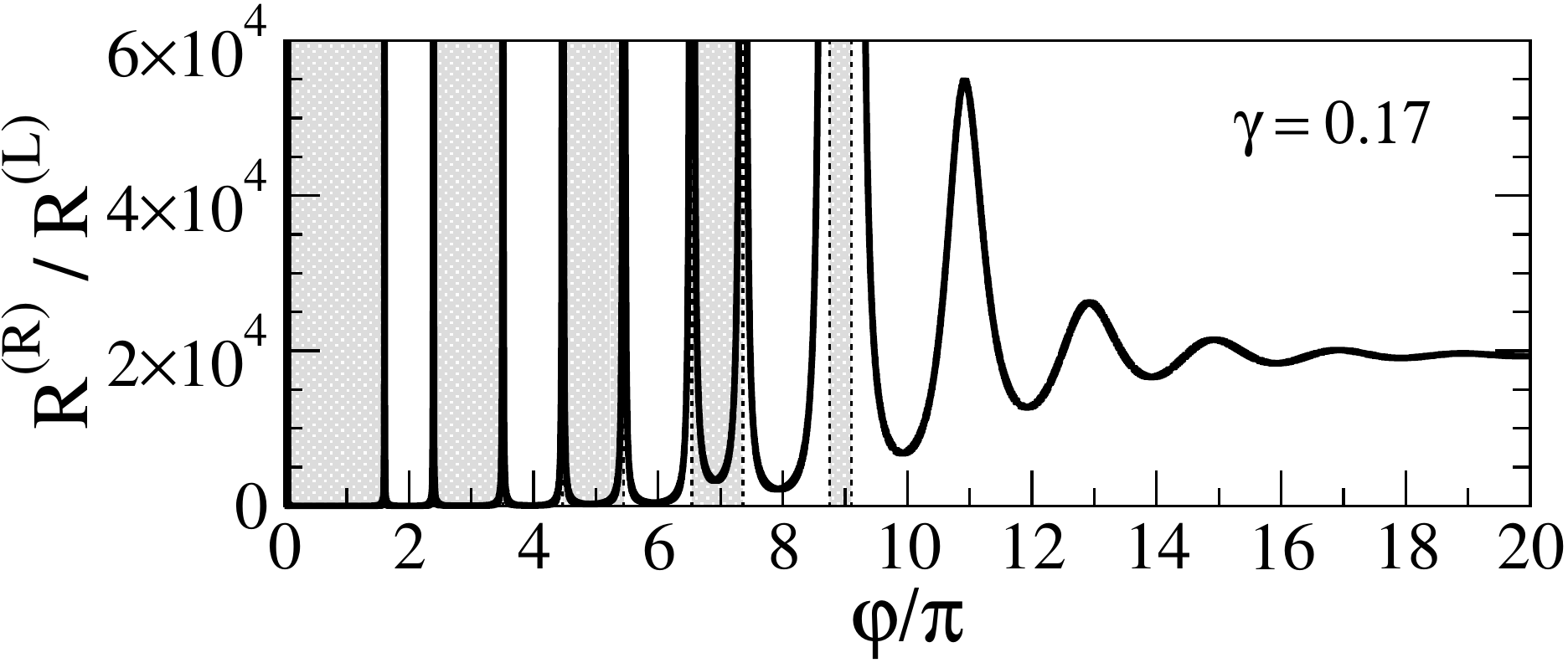}
\caption{Ratio right-to-left reflectances (full curve) versus $\varphi$ for $\gamma=0.17$. Dashed regions are those where $T>1$, see  Fig.~\ref{fig:Fig-03}.}\label{fig:Fig-05}
\end{figure}

The typical behavior of the reflectances \eqref{eq:PMBLG-lR} and \eqref{eq:PMBLG-rR} is depicted in Fig.~\ref{fig:Fig-04}. It demonstrates that the left reflectance is much smaller than the right one. Since at the U-points $\varphi=\varphi_s(\gamma)$ the function $\mathcal{F}(\gamma,\varphi)$ vanishes, the left reflectance $R^{(L)}(\varphi)$ also vanishes, however, the right one, $R^{(R)}(\varphi)$, remains finite,

\begin{eqnarray}
&&R^{(L)}(\varphi_s)=0,\nonumber\\
&&R^{(R)}(\varphi_s)=\frac{4\gamma^2[\cosh(\gamma\varphi_s)-\cos\varphi_s]^2}{(1+\gamma^2)^{2}}\,.\label{eq:PMBLG-R-SpPoints}
\end{eqnarray}
This effect is known as the \emph{unidirectional reflectivity} \cite{Lo11}. It is one of the most important properties of scattering occurring in the $\mathcal{P}\mathcal{T}$-symmetric systems. The ratio between right and left reflectances,
\begin{equation}\label{eq:PMBLG-Rratio}
R^{(R)}/R^{(L)}=\mathcal{F}^2(-\gamma,\varphi)/\mathcal{F}^2(\gamma,\varphi),
\end{equation}
is shown in Fig.~\ref{fig:Fig-05} in dependence on $\varphi$ for $\gamma=0.17$. Note that the right reflectance \eqref{eq:PMBLG-R-SpPoints} exponentially increases with increase of the value of $\varphi_s$, see dashed curve in Fig.~\ref{fig:Fig-04}. Therefore, one can conclude: the higher the value of $\varphi_s$, the stronger the effect of unidirectional reflectivity, see Fig.~\ref{fig:Fig-05}.
%

It is worthwhile to note that the functions $\mathcal{F}(\gamma,\varphi)$ and $\mathcal{F}(-\gamma,\varphi)$ are related by the expression
\begin{eqnarray}\label{eq:PMBLG-F-rel}
&&\mathcal{F}^2(-\gamma,\varphi)-\mathcal{F}^2(\gamma,\varphi)=8[\cosh(\gamma\varphi)-\cos\varphi]\nonumber\\
&&\times[(2+\gamma^2)\sinh(\gamma\varphi)-\gamma\sin\varphi)]\geqslant0
\end{eqnarray}
that directly follows from Eq.~\eqref{eq:F-def}. Since $\gamma>0$ and $\varphi>0$, this relation provides the left reflectance \eqref{eq:PMBLG-lR} to be always smaller than the right one \eqref{eq:PMBLG-rR}, $R^{(L)}<R^{(R)}$. The latter inequality together with Eq.~\eqref{eq:PatTimeRev-TR} gives rise to the condition,
\begin{equation}\label{eq:PMBLG-RT-genRel}
R^{(L)}\leqslant\left|1-T\right|<R^{(R)}.
\end{equation}
The equality at the left part of Eq.~\eqref{eq:PMBLG-RT-genRel} holds only at the U-points $\varphi=\varphi_s(\gamma)$. Remarkably, the condition \eqref{eq:PMBLG-RT-genRel} remains valid for any value of the phase shift $\varphi>0$, as well as for any value $\gamma$ including $\gamma>\gamma_{cr}$.

Within the region $\gamma>\gamma_{cr}$, for a sufficiently large phase shifts $\varphi$ (high frequencies $\omega$) the transmittance \eqref{eq:PMBLG-T} is exponentially small, see Fig.~\ref{fig:Fig-02}~(b). In such a situation the left reflectance \eqref{eq:PMBLG-lR} is larger than $T$, however, smaller than $1$. The right reflectance \eqref{eq:PMBLG-rR} can be shown to exceed unity. Summarizing, one can write, $T<R^{(L)}<1<R^{(R)}$.

For any fixed value of the loss/gain parameter $\gamma$, as $\varphi$ increases and gets into the range where there are no specific points $\varphi_s(\gamma)$ (blank space in Fig.~\ref{fig:Fig-03}), the transmittance exponentially decreases, see Fig.~\ref{fig:Fig-02}. On the other hand,  both the reflectances increase and eventually saturate as is seen in Fig.~\ref{fig:Fig-04},
\begin{eqnarray}\label{eq:PMBLG-Rsat}
&&R^{(L)}(\varphi)\to(1+4\gamma^{-2})^{-1}<1,\nonumber\\
&&R^{(R)}(\varphi)\to1/R^{(L)}(\varphi)>1\quad\mathrm{for}\quad\varphi\to\infty.
\end{eqnarray}
If $\gamma\gg\gamma_{cr}$, the reflectances $R^{(L)}$ and $R^{(R)}$ get the values of unity.

\section{Mismatching leads}

Let us now consider a much more complicated situation when the leads are not perfectly matched with the $(a,b)$ bilayer even in the absence of loss/gain,
\begin{equation}\label{eq:MLBLG-def}
Z_c=\chi Z.
\end{equation}
The new parameter $\chi>0$ specifies the contrast between the impedances $Z_c$ and $Z$, thus determining the mismatch between the $(a,b)$ layers and external leads. Note that the mismatching parameter $\chi$ can be either smaller or greater that one. Now with an inclusion of loss/gain, we have two mechanisms of multiple scattering. The first mechanism is the same as analyzed in previous Section, namely, due to an inclusion of the loss/gain terms. As for the second one, it is the standard one, due to the difference between the impedances $Z_c$ and $Z$. The interplay between these two mechanisms turns out to be highly non-trivial, and is reflected by quite specific transport properties.

Since the parameter $\chi$ is considered to be real, the symmetry \eqref{eq:BLG-Msym} of the total matrix $\hat{M}^{(T)}$ remains the same as in the case of perfect matching. Now the matrix elements read
\begin{eqnarray}\label{eq:MLBLG-Mtot}
M^{(T)}_{11}(\gamma)&=&\frac{\gamma^2\cosh(\gamma\varphi)+\cos\varphi}{1+\gamma^2}+
i\frac{\sin\varphi+\gamma\sinh(\gamma\varphi)}{\chi(1+\gamma^2)}\nonumber\\
&&-\frac{M^{(T)}_{21}(\gamma)-M^{(T)}_{12}(\gamma)}{2},\nonumber\\
M^{(T)}_{12}(\gamma)&=&-\frac{i\mathcal{G}(-\gamma,\chi,\varphi)}{2(1+\gamma^2)},\\
M^{(T)}_{21}(\gamma)&=&\frac{i\mathcal{G}(\gamma,\chi,\varphi)}{2(1+\gamma^2)},\nonumber\\
M^{(T)}_{22}(\gamma)&=&\frac{\gamma^2\cosh(\gamma\varphi)+\cos\varphi}{1+\gamma^2}-i\frac{\sin\varphi+\gamma\sinh(\gamma\varphi)}{\chi(1+\gamma^2)}\nonumber\\[6pt]
&&+\frac{M^{(T)}_{21}(\gamma)-M^{(T)}_{12}(\gamma)}{2}.\nonumber
\end{eqnarray}
Here we have introduced a new function $\mathcal{G}(\gamma,\chi,\varphi)$,
\begin{eqnarray}\label{eq:G-def}
&&\mathcal{G}(\gamma,\chi,\varphi)=\gamma\left[\chi(1+\gamma^2)+\chi^{-1}\right]\sinh(\gamma\varphi)\\
&&-\left[\chi(1+\gamma^2)-\chi^{-1}\right]\sin\varphi+2\gamma\left[\cos\varphi-\cosh(\gamma\varphi)\right],\qquad\nonumber
\end{eqnarray}
which is the generalization of the function $\mathcal{F}(\gamma,\varphi)$ analyzed above. Indeed, for $\chi=1$ we have,
\begin{equation}\label{eq:G-F}
\mathcal{G}(\gamma,\chi=1,\varphi)=\gamma\mathcal{F}(\gamma,\varphi),
\end{equation}
and Eqs.~\eqref{eq:MLBLG-Mtot} are transformed into Eqs.~\eqref{eq:PMBLG-Mtot}.

In accordance with definitions \eqref{eq:T-gen} and \eqref{eq:R-gen}, the analytical expressions for the transmittance $T$, and left and right reflectances, $R^{(L)}$ and $R^{(R)}$, can be expressed in terms of the function $\mathcal{G}(\gamma,\chi,\varphi)$ as follows,
\begin{eqnarray}
&&T=\left[1+\frac{\mathcal{G}(\gamma,\chi,\varphi)\mathcal{G}(-\gamma,\chi,\varphi)}{4(1+\gamma^2)^{2}}\right]^{-1};\label{eq:MLBLG-T}\\
&&\frac{R^{(L)}}{T}=\frac{\mathcal{G}^2(\gamma,\chi,\varphi)}{4(1+\gamma^2)^{2}}\,;\label{eq:MLBLG-lR}\\
&&\frac{R^{(R)}}{T}=\frac{\mathcal{G}^2(-\gamma,\chi,\varphi)}{4(1+\gamma^2)^{2}}\,.\label{eq:MLBLG-rR}
\end{eqnarray}
It is important to stress that for $\chi\neq1$ both functions, $\mathcal{G}(\gamma,\chi,\varphi)$ and $\mathcal{G}(-\gamma,\chi,\varphi)$, can be either positive or negative as functions of the phase shift $\varphi$. This is in contrast with the case of perfect matching considered above for which the function $\mathcal{G}(-\gamma,\chi=1,\varphi)$ is always positive for $\varphi>0$. On the other hand, the symmetry relations \eqref{eq:PatTimeRev-TR} and \eqref{eq:PMBLG-Teven-Rodd} hold true as before.

\section{No loss/gain: Fabry-Perot resonances}

Without the loss/gain ($\gamma=0$) the model can be treated as the Fabry-Perot interferometer with the $(a,b)$ bilayer serving as a single slab of the thickness $d=d_a+d_b$ and with the average refractive index $\bar{n}$,
\begin{equation}\label{eq:av-n}
\bar{n}=\frac{n_a^{(0)}d_a+n_b^{(0)}d_b}{d_a+d_b}=\frac{2n_a^{(0)}d_a}{d}.
\end{equation}
In this case we have,
\begin{equation}\label{eq:G-FP}
\mathcal{G}(\gamma=0,\chi,\varphi)=-\left(\chi-\chi^{-1}\right)\sin\varphi.
\end{equation}
As a consequence, the transfer matrix \eqref{eq:MLBLG-Mtot} meets the time-reversal symmetry \eqref{eq:PatTimeRev}, \eqref{eq:TRSym} with the flow conservation \eqref{eq:FlConsLaw}. Specifically, the transmittance and both reflectances take the form,
\begin{eqnarray}
&&T=\left[1+\frac{1}{4}\left(\chi-\chi^{-1}\right)^2\sin^2\varphi\right]^{-1},\label{eq:T-FP}\\
&&\frac{R}{T}=\frac{R^{(L)}}{T}=\frac{R^{(R)}}{T}=\frac{1}{4}\left(\chi-\chi^{-1}\right)^2\sin^2\varphi.\label{eq:R-FP}
\end{eqnarray}
As one can see, all transmission characteristics, \eqref{eq:T-FP} and \eqref{eq:R-FP}, are periodic functions of the phase shift $\varphi$ with period $\pi$. It should be also noted that for $\gamma=0$ the function $\mathcal{G}$, as well as the transmittance $T$ and the reflectance $R$ have a quite simple symmetry with respect to the mismatching parameter  $\chi$,
\begin{equation}\label{eq:FP-Chi-Sym}
\mathcal{G}(\chi)=-\mathcal{G}(\chi^{-1}),\quad T(\chi)=T(\chi^{-1}),\quad R(\chi)=R(\chi^{-1})
\end{equation}

\begin{figure}[!t]
\includegraphics[width=\columnwidth]{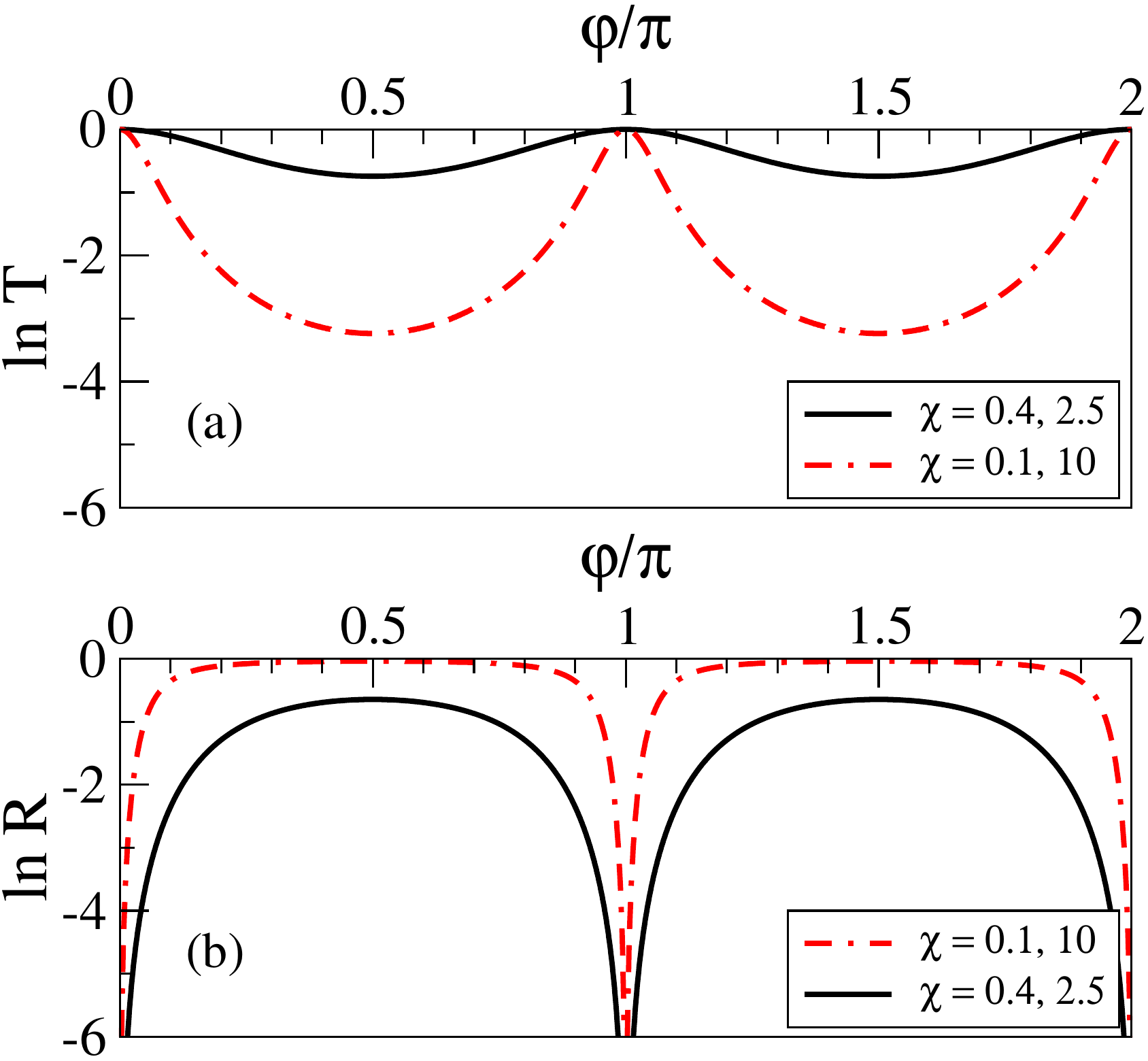}
\caption{Transmittance and reflectance defined by Eqs.~\eqref{eq:T-FP} and \eqref{eq:R-FP} vs phase shift $\varphi$ for few values of $\chi$.}\label{fig:Fig-06}
\end{figure}

The Fabry-Perot resonances are directly associated with the multiple reflections from the interfaces between $(a,b)$ bilayer and mismatching leads $c_L$, $c_R$. The resonance condition emerges when the thickness $d=d_a+d_b$ of the $(a,b)$ bilayer equals an integer multiple of half of the wavelength $\lambda=2\pi c/\omega\bar{n}$ inside the bilayer, i.e., when the phase shift $\varphi$ of the wave passing through the bilayer is multiple to $\pi$,
\begin{equation}\label{eq:FP-res}
\varphi=\varphi_{res}\equiv m\pi,\qquad m=1,2,3,\ldots .
\end{equation}
At the resonances the factor $\sin\varphi$ in Eqs.~\eqref{eq:T-FP}, \eqref{eq:R-FP} vanishes giving rise to the perfect transmission with  $T=1$ and $R=0$, see Eq.~\eqref{eq:PMBLG-TR-gamma0}. Otherwise, $T<1$ and $R=1-T<1$. As a result, the dependencies $T(\varphi)$ and $R(\varphi)$ have an oscillating form as shown in Fig.~\ref{fig:Fig-06}. The amplitude of oscillations is specified by the contrast factor $\left(\chi-\chi^{-1}\right)^2$: The stronger the contrast between $(a,b)$ bilayer and the leads the larger the oscillations.

\section{Loss/gain included: Unidirectional points}

Our analysis shows that when the balanced loss and gain in the layers $a$ and $b$, respectively, are turned on, the perfect transmission, $T=1$, emerges at two different kinds of the U-points, $\varphi_s^{+}$ and $\varphi_s^{-}$ at which either the left or right reflectance vanishes. This happens due to vanishing either the function $\mathcal{G(\gamma,\chi,\varphi)}$ or the function $\mathcal{G(-\gamma,\chi,\varphi)}$, see Eqs.~\eqref{eq:MLBLG-T} -- \eqref{eq:MLBLG-rR},
\begin{equation}\label{eq:MLBLG-SpRes}
\mathcal{G}(\pm\gamma,\chi,\varphi)=0\,\to\,\varphi=\varphi_s^{\pm}(\gamma,\chi)\,\to\,T(\varphi_s^{\pm})=1.
\end{equation}

\begin{figure}[t]
\centering
\includegraphics[width=\columnwidth]{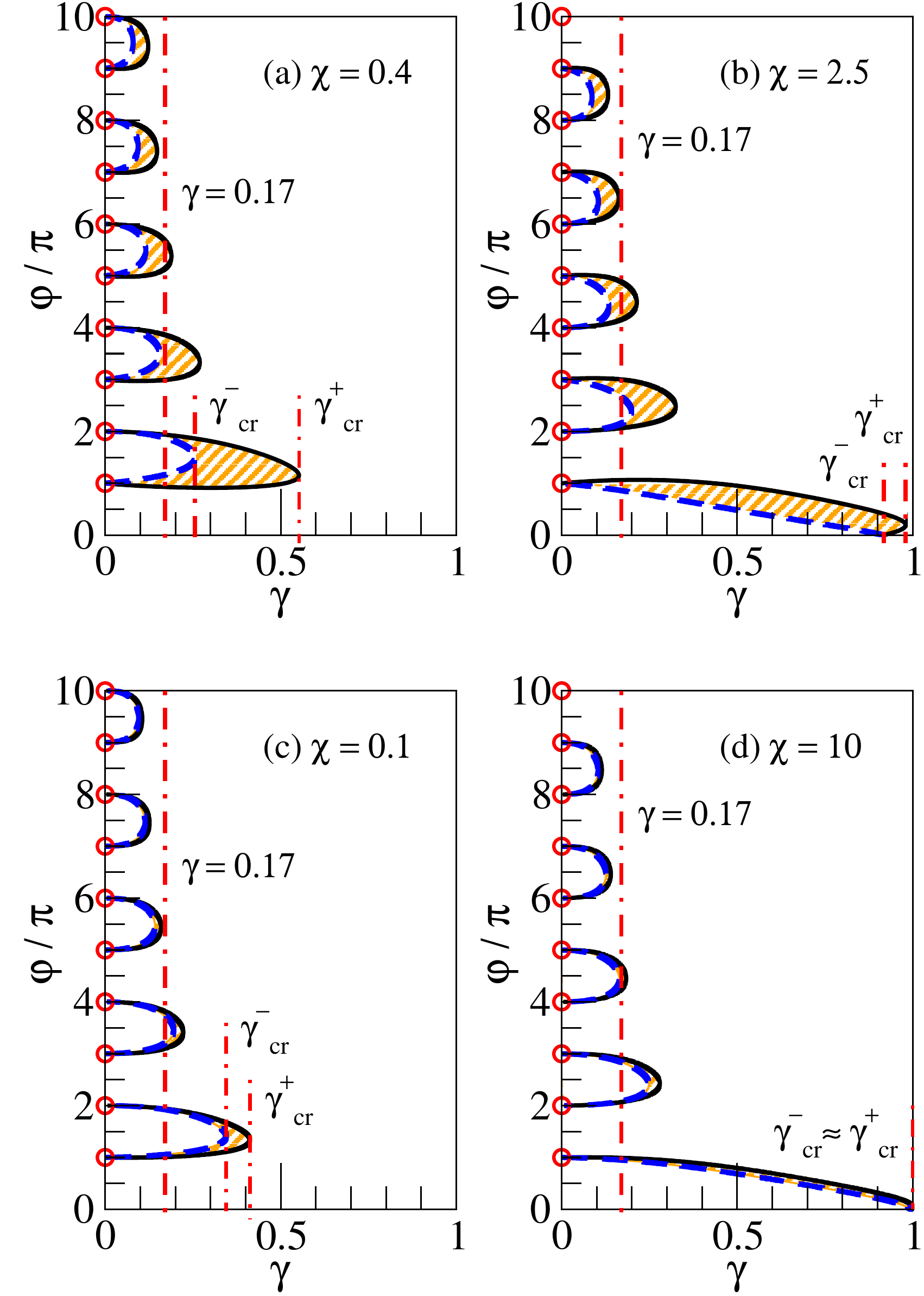}
\caption{Relation between $\gamma$ and $\varphi/\pi$ determining the U-points, for few values of the parameter $\chi$. Solid curves correspond to the U-points $\varphi=\varphi^{+}_s(\gamma,\chi)$ and dashed curves - to the U-points $\varphi=\varphi^{-}_s(\gamma,\chi)$. Along these curves the transmission is perfect, $T=1$, and one of the reflectances vanishes. Inside shaded regions the transmission is larger than one, $T(\varphi)>1$, whereas $T(\varphi)<1$ outside these regions. Circles stand for the Fabry-Perot resonances, $\varphi=\varphi_{res}$, see Eq.~\eqref{eq:FP-res}. Vertical dash-dotted lines are shown for $\gamma=0.17$ and for the thresholds $\gamma_{cr}^{\pm}(\chi)$.}\label{fig:Fig-07}
\end{figure}

Fig.~\ref{fig:Fig-07} presents four examples of the dependence of the U-points $\varphi_s^{\pm}(\gamma,\chi)$ on the loss/gain parameter $\gamma$ for given values of the mismatching parameter $\chi$. One can see that this dependence is quite sophisticated and very sensitive to the value of $\chi$. In particular, in the presence of balanced loss/gain the symmetry relation \eqref{eq:FP-Chi-Sym} turns out to be broken. Indeed, the curves in Figs.~\ref{fig:Fig-07}~(a) and (c) drastically differ from those in Figs.~\ref{fig:Fig-07}~(b) and (d), respectively, despite the fact that the values of mismatching $\chi=0.4$, $\chi=2.5$ in Figs.~\ref{fig:Fig-07}~(a),~(b) and $\chi=0.1$, $\chi=10$ in Figs.~\ref{fig:Fig-07}~(c),~(d) are mutually inverses.

\begin{figure}[t]
\centering
\includegraphics[width=\columnwidth]{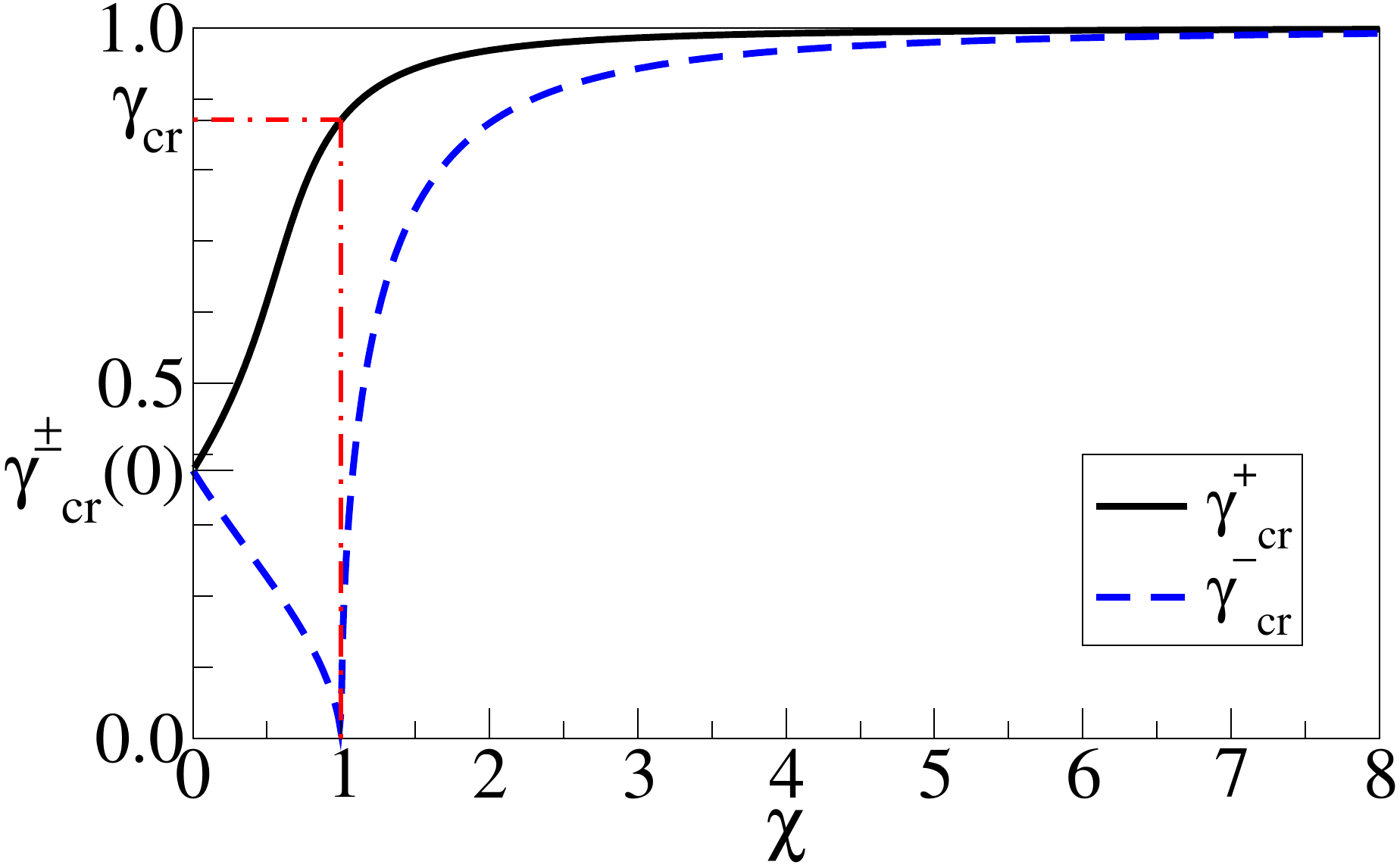}
\caption{Threshold values $\gamma_{cr}^{+}(\chi)$ (solid curve) and $\gamma_{cr}^{-}(\chi)$ (dashed curve) versus the mismatching parameter $\chi$.}\label{fig:Fig-08}
\end{figure}

One should emphasize that equations \eqref{eq:MLBLG-SpRes} have the solutions $\varphi=\varphi_s^{+}(\gamma,\chi)$ and $\varphi=\varphi_s^{-}(\gamma,\chi)$ only when the loss/gain parameter $\gamma$ does not exceed its threshold (critical) values
$\gamma_{cr}^{+}(\chi)$ and $\gamma_{cr}^{-}(\chi)$, respectively,
\begin{subequations}\label{eq:MLBLG-gamma-thresh-def}
\begin{eqnarray}
&&\varphi=\varphi_s^{+}(\gamma,\chi)\quad\mathrm{exists\ if}\quad\gamma\leqslant\gamma_{cr}^{+}(\chi),\\
&&\varphi=\varphi_s^{-}(\gamma,\chi)\quad\mathrm{exists\ if}\quad\gamma\leqslant\gamma_{cr}^{-}(\chi).
\end{eqnarray}
\end{subequations}
From Fig.~\ref{fig:Fig-07} one can readily conclude that these two thresholds meet the conditions
\begin{equation}\label{eq:MLBLG-gamma-thresh-rel}
0<\gamma_{cr}^{-}(\chi)<\gamma_{cr}^{+}(\chi)<1
\end{equation}
for any finite value of $\chi$. Then, as one can realize from Eq.~\eqref{eq:G-F}, in the absence of the contrast ($\chi=1$) the lower threshold $\gamma_{cr}^{-}(\chi)$ vanishes, whereas the upper one $\gamma_{cr}^{+}(\chi)$ coincides with the critical value $\gamma_{cr}$ inherent for the U-points $\varphi_s(\gamma)$, see previous Section. For a strong contrast (when the parameter $\chi$ is either very small or very large) both thresholds $\gamma_{cr}^{\pm}(\chi)$ approach the same limits, that are, however, different for $\chi\to0$ and $\chi\to\infty$. Summarizing, we can write
\begin{subequations}\label{eq:MLBLG-gamma-thresh-lim}
\begin{eqnarray}
&&\gamma_{cr}^{-}(1)=0,\qquad\gamma_{cr}^{+}(1)=\gamma_{cr}\approx0.8703,\\
&&\gamma_{cr}^{\pm}(0)\approx0.377,\qquad\gamma_{cr}^{\pm}(\infty)=1.
\end{eqnarray}
\end{subequations}
Note that the thresholds $\gamma_{cr}^{\pm}(\chi)$ reveal different behavior in dependence on the parameter $\chi$ as is shown in Fig.~\ref{fig:Fig-08}. While $\gamma_{cr}^{+}(\chi)$ monotonously increases from the initial value $\gamma_{cr}^{\pm}(0)\approx0.377$ to $\gamma_{cr}^{\pm}(\infty)=1$, the threshold $\gamma_{cr}^{-}(\chi)$ is non-monotonic function of $\chi$. It decreases within the interval $(0,1)$ from the same initial value $\gamma_{cr}^{\pm}(0)\approx0.377$ up to zero and only after begins to increase approaching the limit $\gamma_{cr}^{\pm}(\infty)=1$.

Due to existence of the U-points $\varphi_s^{\pm}(\gamma,\chi)$, in Fig.~\ref{fig:Fig-07} the shaded regions of an anomalous transmission with $T>1$ emerge. These regions are located between $\varphi=\varphi_s^{-}(\gamma,\chi)$ (dashed curves) and $\varphi=\varphi_s^{+}(\gamma,\chi)$ (solid curves) for the values of $\gamma$ from the interval $0<\gamma<\gamma_{cr}^{+}(\chi)$. Fig.~\ref{fig:Fig-07} displays that for $\chi\neq1$ the smaller the parameter $\gamma$, the larger the number of anomalous regions, however, the more narrow their phase shift (frequency) range. This fact is in contradiction to the case of perfect matching ($\chi=1$) for which the range of anomalous regions increases with a decrease of $\gamma$, see Fig.~\ref{fig:Fig-03}. Since both functions $\mathcal{G}(\pm\gamma,\chi,\varphi)$ become positive-valued for $\gamma>\gamma_{cr}^{+}(\chi)$, there are no U-points and the transmittance \eqref{eq:MLBLG-T} is always smaller than one, $T<1$.

In accordance with Eqs.~\eqref{eq:G-F} and \eqref{eq:PMBLG-SpPoints}, in the case of perfect matching ($\chi=1$), the U-points $\varphi_s^{+}(\gamma,\chi)$ transform onto those $\varphi_s(\gamma)$ considered in previous Section. At the same time, the U-points, $\varphi_s^-(\gamma,\chi)$, having the zero-threshold $\gamma_{cr}^{-}(\chi=1)=0$, disappears. Thus,
\begin{subequations}\label{eq:MLBLG-phi-matching}
\begin{eqnarray}
&&\varphi_s^{+}(\gamma,\chi=1)=\varphi_s(\gamma),\\
&&\varphi_s^{-}(\gamma,\chi=1)\quad\mathrm{does\ not\ exist}.
\end{eqnarray}
\end{subequations}

On the other hand, when $\gamma\to0$ and $\chi\neq1$, both U-points $\varphi=\varphi_s^{\pm}(\gamma,\chi)$ degenerate into the Fabry-Perot resonances \eqref{eq:FP-res} denoted by circles in Fig.~\ref{fig:Fig-07},
\begin{equation}
\varphi_s^{\pm}(\gamma=0,\chi)=\varphi_{res}.
\end{equation}
Therefore, the balanced loss/gain splits the Fabry-Perot resonances into the U-points, giving rise anomalous regions with $T>1$. Remarkably, when the contrast is strong, i.e. $|\chi-\chi^{-1}|\gg1$, the splitting is extremely weak. Correspondingly, the phase shift range of the anomalous regions with $T>1$ is also extremely small.

\begin{figure}[t]
\centering
\includegraphics[width=\columnwidth]{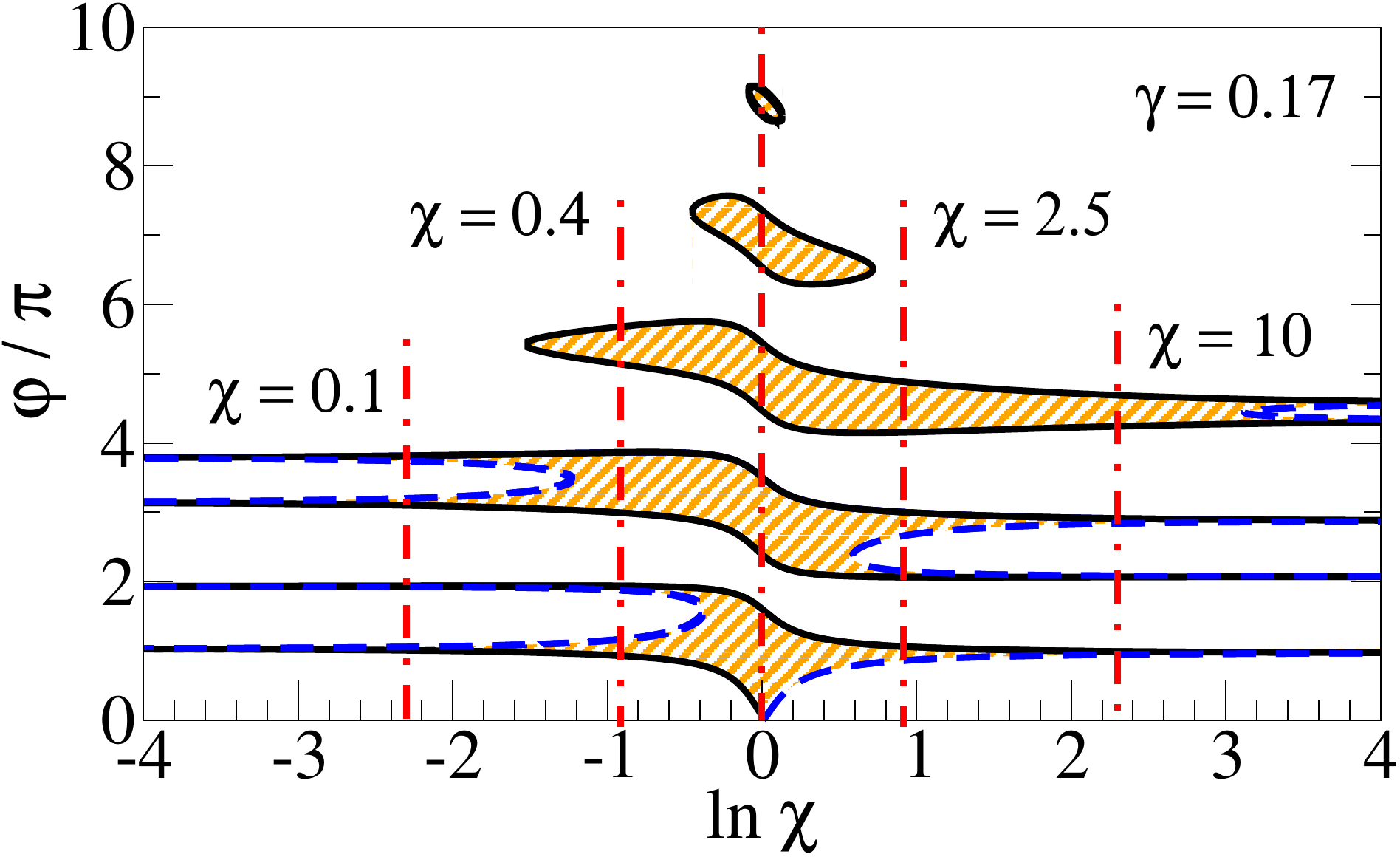}
\caption{Relation between $\varphi/\pi$ and $\chi$ for $\gamma=0.17$. Solid curves correspond to the U-points $\varphi=\varphi^{+}_s(\gamma,\chi)$ and dashed curves - to the U-points $\varphi=\varphi^{-}_s(\gamma,\chi)$. Along these curves the transmission is perfect, $T=1$, and one of the reflectances vanishes. Inside shaded regions the transmission is larger than one, $T(\varphi)>1$, while $T(\varphi)<1$ outside them. Vertical dash-dotted lines stand for the values of $\chi$ used in Fig.~\ref{fig:Fig-07}.}\label{fig:Fig-09}
\end{figure}

Fig.~\ref{fig:Fig-09} depicts a quite sophisticated (however, typical) dependence of the U-points, $\varphi=\varphi^{+}_s(\gamma,\chi)$ (solid curves) and $\varphi=\varphi^{-}_s(\gamma,\chi)$ (dashed curves), on the mismatching parameter $\chi$ for a given value of the loss/gain parameter $\gamma$. These functions are asymmetric with respect to the axis $\chi=1$, i.e. the symmetry \eqref{eq:FP-Chi-Sym} is broken due to the balanced loss/gain as it was noted above. Then, in line with Eqs.~\eqref{eq:MLBLG-phi-matching}, all curves $\varphi=\varphi^{-}_s(\gamma,\chi)$ never approach the axis $\chi=1$, whereas all curves $\varphi=\varphi^{+}_s(\gamma,\chi)$ always cross it with an increase of $\chi$.

The number of the U-points $\varphi_s^{\pm}(\gamma,\chi)$ is finite and depends on the values of $\gamma$ and $\chi$. At fixed value $\chi$, the smaller the parameter $\gamma$, the larger the number of these points, see Fig.~\ref{fig:Fig-07}. When we fix the parameter $\gamma$ and increase the contrast factor $|\chi-\chi^{-1}|$, the number of U-points $\varphi=\varphi^{+}_s(\gamma,\chi)$ decreases, however, the number of U-points $\varphi=\varphi^{-}_s(\gamma,\chi)$ increases approaching the same limiting value that does not depend on $\chi$. This happens since the curves of the U-points, $\varphi=\varphi^{+}_s(\gamma,\chi)$ and $\varphi=\varphi^{-}_s(\gamma,\chi)$, come closer to each other as the contrast factor $|\chi-\chi^{-1}|$ becomes larger, see Fig.~\ref{fig:Fig-09}.

Let us now examine the case of strong contrast between the unperturbed impedance $Z$ of the $(a,b)$ bilayer and impedance $Z_c$ of the external leads,
\begin{equation}\label{eq:StrongContrast-def}
|\chi-\chi^{-1}|\gg1.
\end{equation}
This case can be realized when the mismatching parameter $\chi$ is either very small ($\chi\ll1$) or very large ($\chi\gg1$) and where the characteristic function \eqref{eq:G-def} follows the asymptotics,
\begin{subequations}\label{eq:G-StrContr}
\begin{eqnarray}
&&\mathcal{G}(\gamma,\chi,\varphi)\approx\chi^{-1}[\gamma\sinh(\gamma\varphi)+\sin\varphi]\nonumber\\
&&\qquad\mathrm{for}\quad\chi\ll1,\label{eq:G-ChiZero}\\[6pt]
&&\mathcal{G}(\gamma,\chi,\varphi)\approx\chi(1+\gamma^2)[\gamma\sinh(\gamma\varphi)-\sin\varphi]\nonumber\\
&&\qquad\mathrm{for}\quad\chi\gg1.\label{eq:G-ChiInfinite}
\end{eqnarray}
\end{subequations}
Due to the evenness of both asymptotics \eqref{eq:G-StrContr} with respect to $\gamma$, the curves of the U-points $\varphi_s^{+}(\gamma,\chi)$ and $\varphi_s^{-}(\gamma,\chi)$ approach each other in accordance with Fig.~\ref{fig:Fig-09}, having the same thresholds, $\gamma_{cr}^{\pm}(0)\approx0.377$ for $\chi\ll1$ and $\gamma_{cr}^{\pm}(\infty)=1$ for $\chi\gg1$, see Eqs.~\eqref{eq:MLBLG-gamma-thresh-lim}. Remarkably, for a strong contrast the limiting values of the U-points $\varphi_s^{\pm}(\gamma,\chi=0)$ and $\varphi_s^{\pm}(\gamma,\chi=\infty)$, defined, respectively, by the zeros of Eqs.~\eqref{eq:G-ChiZero} and \eqref{eq:G-ChiInfinite}, depend only on the parameter $\gamma$. It is clear that the asymptotics \eqref{eq:G-StrContr} cannot describe the anomalous regions with $T>1$, since these regions are extremely small in the case of the strong contrast. In order to take them into account, Eqs.~\eqref{eq:G-StrContr} have to be corrected with the terms of the next order in $\chi$ or $\chi^{-1}$.

\section{Unidirectional reflectivity}

For a strong contrast \eqref{eq:StrongContrast-def} and apart from the anomalous regions where $T>1$, the general expressions \eqref{eq:MLBLG-T} and \eqref{eq:MLBLG-lR}, \eqref{eq:MLBLG-rR} for the transmittance and both reflectances can be approximated as
\begin{eqnarray}
&&T\approx\left[1+\frac{\chi^{-2}}{4(1+\gamma^2)^2}\left[\gamma\sinh(\gamma\varphi)+\sin\varphi\right]^2\right]^{-1},\label{eq:T-ChiZero}\\
&&\frac{R}{T}\approx\frac{R^{(L)}}{T}\approx\frac{R^{(R)}}{T}\approx
\frac{\chi^{-2}\left[\gamma\sinh(\gamma\varphi)+\sin\varphi\right]^2}{4(1+\gamma^2)^2}\label{eq:R-ChiZero}\qquad\\
&&\qquad\mathrm{for}\quad\chi\ll1;\nonumber
\end{eqnarray}
and
\begin{eqnarray}
&&T\approx\left[1+\frac{\chi^2}{4}\left[\gamma\sinh(\gamma\varphi)-\sin\varphi\right]^2\right]^{-1},\label{eq:T-ChiInfinity}\\
&&\frac{R}{T}\approx\frac{R^{(L)}}{T}\approx\frac{R^{(R)}}{T}\approx\frac{\chi^2}{4}\left[\gamma\sinh(\gamma\varphi)-\sin\varphi\right]^2
\qquad\label{eq:R-ChiInfinity}\\
&&\qquad\mathrm{for}\quad\chi\gg1.\nonumber
\end{eqnarray}
The main conclusion that follows from these estimates is that both left and right reflectances coincide. As a consequence, the flow conservation law \eqref{eq:FlConsLaw} is restored, similarly to what happens in the case of no loss/gain ($\gamma=0$), see Eqs.~\eqref{eq:G-FP} -- \eqref{eq:R-FP}, and in spite of the fact that the symmetry condition \eqref{eq:FP-Chi-Sym} remains to be broken. However, it is worthwhile to note that for $\gamma=0$ the obtained asymptotics \eqref{eq:G-StrContr} -- \eqref{eq:R-ChiInfinity} coincide with those resulting from the corresponding expressions \eqref{eq:G-FP}, \eqref{eq:T-FP} and \eqref{eq:R-FP} when either $\chi\ll1$ or $\chi\gg1$.

\begin{figure}[!ht]
\includegraphics[width=\columnwidth]{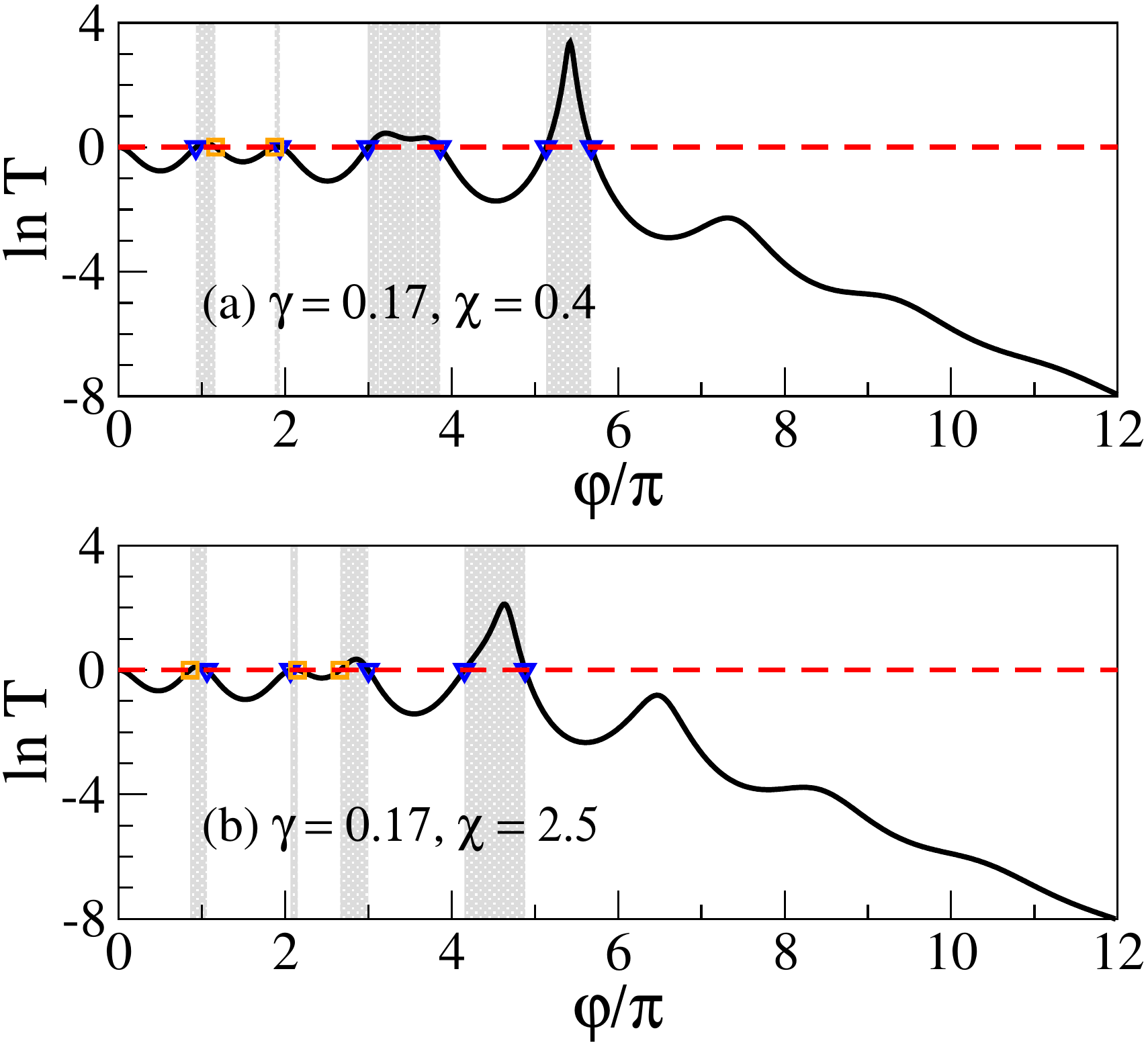}
\caption{The logarithm of $T$ defined by Eq.~\eqref{eq:MLBLG-T} vs phase shift $\varphi$. Shaded regions correspond to $T>1$, small triangles and squares stand for the U-points $\varphi_s^{+}(\gamma,\chi)$ and $\varphi_s^{-}(\gamma,\chi)$, respectively.}\label{fig:Fig-10}
\end{figure}

Fig.~\ref{fig:Fig-10} exhibits two typical examples of the dependence $T(\varphi)$, see Eq.~\eqref{eq:MLBLG-T}, for the loss/gain parameter $0<\gamma<\gamma_{cr}^{-}(\chi)<\gamma_{cr}^{+}(\chi)$ and two values of the parameter $\chi<1$ and $\chi>1$. As discussed above, at the U-points  the transmission is perfect, $T(\varphi_s^{\pm})=1$. One can detect that for $\chi=0.4$ ($\chi=2.5$) there are six (five) U-points $\varphi_s^{+}(\gamma,\chi)$ and two (three) U-points $\varphi_s^{-}(\gamma,\chi)$, in accordance with Figs.~\ref{fig:Fig-07}~(a),~(b) and Fig.~\ref{fig:Fig-09}. When the phase-shift $\varphi$ increases and crosses the largest U-point $\varphi_s^{+}(\gamma,\chi)$, the transmittance exponentially decreases.

\begin{figure}[!ht]
\centering
\includegraphics[width=\columnwidth]{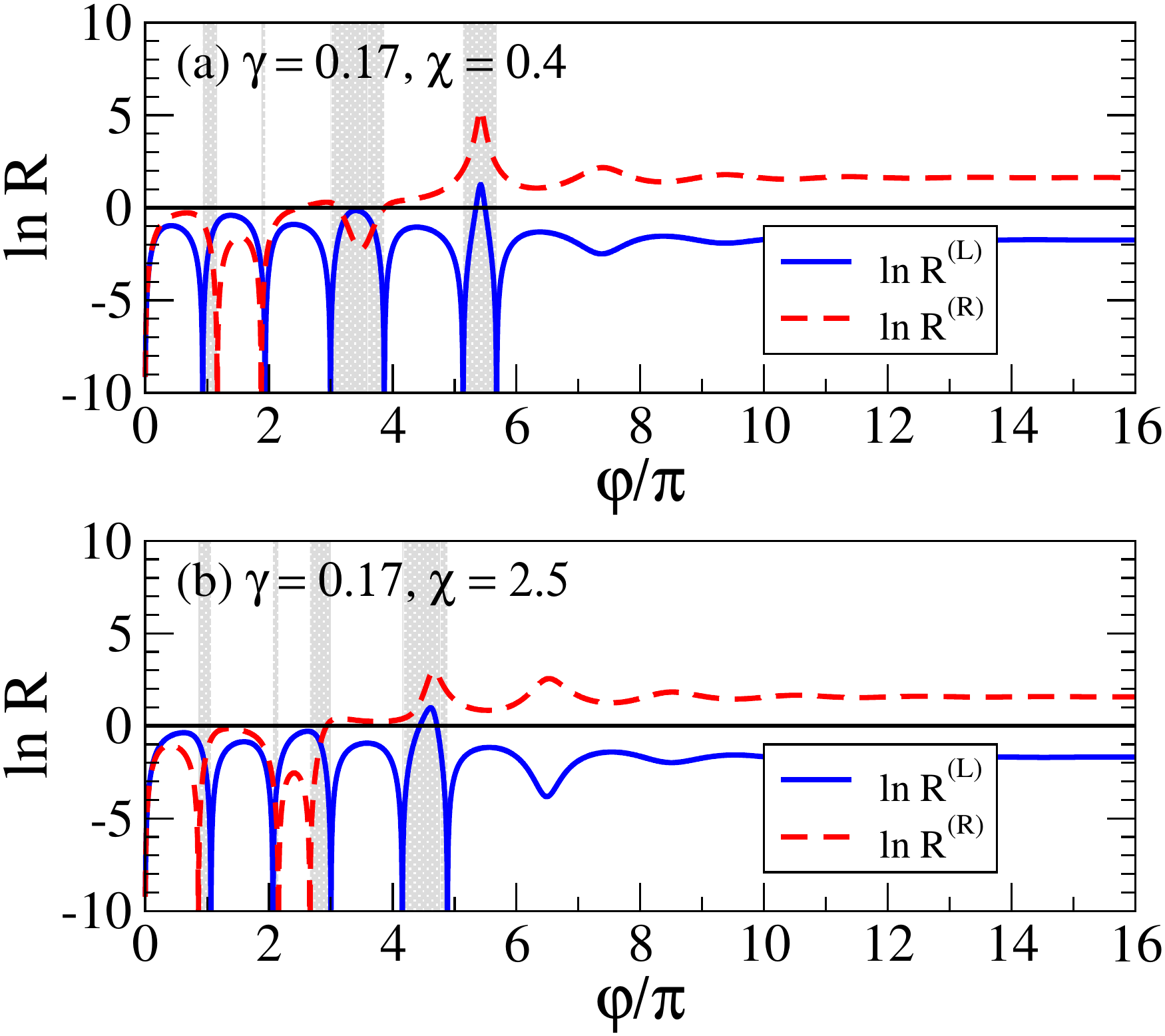}
\caption{The logarithms of left $R^{(L)}$ and right $R^{(R)}$ reflectances vs the phase shift $\varphi$ for $\gamma=0.17$ and two values of $\chi$.}\label{fig:Fig-11}
\end{figure}

When the value of the loss/gain parameter $\gamma$ is located within the interval $0<\gamma<\gamma_{cr}^{-}(\chi)<\gamma_{cr}^{+}(\chi)$, a highly non-trivial behavior is displayed by the left/right reflectances \eqref{eq:MLBLG-lR} and \eqref{eq:MLBLG-rR}, see Fig.~\ref{fig:Fig-11}. Since the function $\mathcal{G}(\gamma,\chi,\varphi)$ vanishes at the U-points $\varphi=\varphi^{+}_s(\gamma,\chi)$, the left reflectance $R^{(L)}(\varphi)$ also vanishes, however, the right one, $R^{(R)}(\varphi)$, remains finite,
\begin{eqnarray}
&&R^{(L)}(\varphi^{+}_s)=0,\label{eq:MLBLG-R-SpPlusPoints}\\
&&R^{(R)}(\varphi^{+}_s)=\frac{4\gamma^2[\cosh(\gamma\varphi^{+}_s)-\cos\varphi^{+}_s]^2}{(1+\gamma^2)^{2}}\,.\nonumber
\end{eqnarray}
On the contrary, for another type of the U-points, $\varphi=\varphi^{-}_s(\gamma,\chi)$, the other function $\mathcal{G}(-\gamma,\chi,\varphi)$ vanishes. Therefore, the right reflectance $R^{(R)}(\varphi)$ vanishes instead of the left one, $R^{(L)}(\varphi)$,
\begin{eqnarray}
&&R^{(L)}(\varphi^{-}_s)=\frac{4\gamma^2[\cosh(\gamma\varphi^{-}_s)-\cos\varphi^{-}_s]^2}{(1+\gamma^2)^{2}},\nonumber\\
&&R^{(R)}(\varphi^{-}_s)=0\,.\label{eq:MLBLG-R-SpMinusPoints}
\end{eqnarray}
Thus, as the phase shift $\varphi$ changes the right/left unidirectional reflectivity can be switched to the opposite, left/right, one. Remarkably, such a phenomenon of double unidirectional reflectivity is originated from the mismatching between the bilayer optical pattern and the connecting leads ($\chi\neq1$). If the matching becomes perfect ($\chi=1$) the left unidirectional reflectivity \eqref{eq:MLBLG-R-SpMinusPoints} disappears and we arrive at the situation studied in the previous Section, see Eqs.~\eqref{eq:PMBLG-R-SpPoints}.

By comparing Eqs.~\eqref{eq:PMBLG-R-SpPoints}, with \eqref{eq:MLBLG-R-SpPlusPoints} and \eqref{eq:MLBLG-R-SpMinusPoints}, one can see that they have a quite similar form. The only, however crucial, difference is in the form of the expressions defining the U-points $\varphi_s(\gamma)$ and $\varphi^{\pm}_s(\gamma,\chi)$ that should be substituted when passing from the case with $\chi=1$ to that of $\chi\neq 1$. This generalization shows how the U-points $\varphi^{\pm}_s(\gamma,\chi)$ depend on the mismatching $\chi$, see also Eqs.~\eqref{eq:MLBLG-phi-matching}.

It is worthwhile to note that the squared functions $\mathcal{G}(\gamma,\chi,\varphi)$ and $\mathcal{G}(-\gamma,\chi,\varphi)$ are connected by the relation,
\begin{eqnarray}\label{eq:MLBLG-G-rel}
&&\mathcal{G}^2(-\gamma,\chi,\varphi)-\mathcal{G}^2(\gamma,\chi,\varphi)\nonumber\\
&&=4[\cosh(\gamma\varphi)-\cos\varphi]\big[\mathcal{G}(-\gamma,\chi,\varphi)+\mathcal{G}(\gamma,\chi,\varphi)\big]\nonumber\\
&&=8[\cosh(\gamma\varphi)-\cos\varphi]\big[\chi(1+\gamma^2)(\gamma\sinh(\gamma\varphi)-\sin\varphi)\nonumber\\
&&+\chi^{-1}(\gamma\sinh(\gamma\varphi)+\sin\varphi)\big],
\end{eqnarray}
that can be derived by the direct use of definition \eqref{eq:G-def}. If $\gamma>\gamma_{cr}^{+}(\chi)$, there are no U-points since both $\mathcal{G}(-\gamma,\chi,\varphi)$ and $\mathcal{G}(\gamma,\chi,\varphi)$ become the positive-valued functions. This difference is always positive providing the left reflectance \eqref{eq:MLBLG-lR} to be smaller than the right one \eqref{eq:MLBLG-rR}, $R^{(L)}(\varphi)<R^{(R)}(\varphi)$. Taking into account the equality \eqref{eq:PatTimeRev-TR} and the fact that for the same values of $\gamma$ the transmittance \eqref{eq:MLBLG-T} is always smaller than one, $T(\varphi)<1$, we can readily come to the conditions
\begin{equation}\label{eq:MLBLG-RT-parRel}
R^{(L)}<1-T<R^{(R)}\qquad\mathrm{for}\quad\gamma>\gamma_{cr}^{+}(\chi).
\end{equation}
Remarkably, these conditions are valid for any value of the phase shift $\varphi>0$ and any value of the mismatching parameter $\chi>0$.

Within the region $\gamma>\gamma_{cr}^{+}(\chi)$ and for a sufficiently large phase shifts $\varphi$ (high frequencies $\omega$) the transmittance \eqref{eq:MLBLG-T} becomes exponentially small. In such a situation the left reflectance \eqref{eq:MLBLG-lR} turns out to be larger than $T$, however smaller than $1$. As a consequence, the relation \eqref{eq:MLBLG-RT-parRel} is reduced to $T<R^{(L)}<1<R^{(R)}$.

For any value of $\gamma$, with an increase of the phase shift $\varphi$ beyond the value determining the absence of the U-points $\varphi^{\pm}_s(\gamma,\chi)$ (see blank space in Fig.~\ref{fig:Fig-07}), the transmittance exponentially decreases as shown in Fig.~\ref{fig:Fig-10}, whereas both reflectances increase and eventually saturate, see Fig.~\ref{fig:Fig-11},
\begin{eqnarray}\label{eq:MLBLG-Rsat}
&&R^{(L)}(\varphi)\to\frac{\chi+(\chi+\chi^{-1}-2)\gamma^{-2}}{\chi+(\chi+\chi^{-1}+2)\gamma^{-2}}<1,\nonumber\\
&&R^{(R)}(\varphi)\to1/R^{(L)}(\varphi)>1\quad\mathrm{for}\quad\varphi\to\infty.
\end{eqnarray}
In the case of perfect matching ($\chi=1$) the asymptotes \eqref{eq:MLBLG-Rsat} coincide with those defined by Eqs.~\eqref{eq:PMBLG-Rsat}.

\section{Conclusions}
In conclusion, we have studied an optical bilayer model with balanced loss/gain, paying the main attention to the effect of unidirectional reflectivity. The model is formally not the $\mathcal{P}\mathcal{T}$-symmetric one, however, its transport properties have much in common with those emerging in the $\mathcal{P}\mathcal{T}$-symmetric models. The analytical analysis demonstrates that an inclusion of the mismatching between the scattering part and perfect leads, results in a quite unexpected effect termed "double-sided unidirectional reflectivity".

We have found that in addition to the well known effect of vanishing of the reflectance from one side of the scattering setup, another set of specific values of frequency arises at which the reflectance from the opposite side vanishes as well. We term the values of the frequency at which either left or right reflectance vanishes as the U-points, in order to stress these points are, in general, different from the "exceptional points" known to be related to the unidirectional reflectivity in some of the $\mathcal{P}\mathcal{T}$-symmetric models. Thus, when changing the frequency, one can observe the switching between the two types of zero reflectivity, and such a switching can be very sensitive to the frequency value.

Our analytical approach allows one to identify the conditions under which both types of the U-points emerge, and fully describe the properties of the transmission and reflection of the electromagnetic waves in dependence on the model parameters. The analytical expressions are complemented by the numerical data. The obtained results can be helpful for the experimental implementation of an anomalous transport in optic devices, for example, for the creation of optical switching devices with a high sensitivity to the frequency of scattering waves.

\begin{acknowledgments}
The authors are thankful to Prof. D.~Christodoulides for fruitful discussions. This work was supported by the CONACYT (M\'exico).
\end{acknowledgments}



\end{document}